\documentclass[fleqn,usenatbib]{mnras}

\usepackage{newtxtext,newtxmath}
\usepackage[english]{babel}

\usepackage[T1]{fontenc}

\DeclareRobustCommand{\VAN}[3]{#2}
\let\VANthebibliography\thebibliography
\def\thebibliography{\DeclareRobustCommand{\VAN}[3]{##3}\VANthebibliography}

\usepackage{graphicx}	
\usepackage{amsmath}	
\usepackage{xcolor}
\definecolor{falured}{rgb}{0.5, 0.09, 0.09}

\usepackage{multirow}

\title[VLBI Imaging in L2 Orbit]{High Resolution Imaging of a Black Hole Shadow with Millimetron Orbit around Lagrange Point L2}

\author[S. F. Likhachev et al.]{S.~F.~Likhachev$^{1}$, A.~G.~Rudnitskiy$^{1}$\thanks{E-mail: arud@asc.rssi.ru}, M.~A.~Shchurov$^{1}$, A.~S.~Andrianov$^{1}$, A.~M.~Baryshev$^{2}$, 
\newauthor S.~V.~Chernov$^{1}$, V.~I.~Kostenko$^{1}$\\ 
$^{1}$Astro Space Center, Lebedev Physical Institute, Russian Academy of Sciences, Profsoyuznaya str. 84/32, Moscow, 117997, Russia\\
$^{2}$Kapteyn Astronomical Institute, University of Groningen, P.O. Box 800, 9700 AV Groningen, the Netherlands
}
\date{Accepted XXX. Received YYY; in original form ZZZ}

\pubyear{2021}

\begin{document}
\label{firstpage}
\pagerange{\pageref{firstpage}--\pageref{lastpage}}
\maketitle

\begin{abstract}
Imaging of the shadow around supermassive black hole (SMBH) horizon with a very long baseline interferometry (VLBI) is recognized recently as a powerful tool for experimental testing of Einstein's General relativity. The Event Horizon Telescope (EHT) has demonstrated that an Earth-extended VLBI with the maximum long base ($D=10,700$ km) can provide a sufficient angular resolution $\theta\sim 20~\mu$as at $\lambda=1.3$ mm ($\nu=230$ GHz) for imaging the shadow around SMBH located in the galaxy M87$^\ast$. However, the accuracy of critically important characteristics, such as the asymmetry of the crescent-shaped bright structure around the shadow and the sharpness of a transition zone between the shadow floor and the bright crescent silhouette, both of order $\Delta\theta\sim 4~\mu$as, is still to be improved. In our previous paper we have shown that Space-Earth VLBI observation within a joint Millimetron and EHT configuration at the near-Earth high elliptical orbit (HEO) can considerably improve the image quality. Even more solid grounds for firm experimental validation of General relativity can be obtained with a higher resolution available within the joint Millimetron and EHT program at the Lagrangian point L2 in the Sun-Earth system with an expected imaging resolution at 230~GHz of $\Delta\theta\sim 5~\mu$as. In this paper we argue that in spite of limitations of L2 orbit, an adequate sparse $(u,v)$ coverage can be achieved and the imaging of the shadows around Sgr~A$^\ast$ and M87$^\ast$ can be performed with a reasonable quality.
\end{abstract}



\section{Introduction}

Observations of supermassive black holes (SMBH) in far-infrared (FIR) and sub-millimeter wavebands with high angular resolution provide a unique opportunity to test the Einstein's theory of General relativity \citep[GR, see discussion in][and references therein]{Cunha2018,2019ApJ...875L...1E,Psaltis2019, Berti2019}. Direct observations of SMBH, the objects that by definition are unseen, are possible through imaging of a bright crescent-like ring around the black hole.

The shadow diameter is $d_{s} \approx 5.2 R_{s}$ for the non spinning black hole, while the minimum diameter is $\approx4.84 R_{s}$ ($R_s=2GM/c^2$ is the radius of Schwarzschild surface) for the maximum spinning black hole seen along with the axis of the spin \citep{Chan2013}. The shadow shape is determined by the physical characteristics of the SMBH itself (its mass $M$ and spin $a$) and by the angle between the black hole spin and the observer line of sight. Observations of the SMBH shadow and brightness distribution in the crescent provide the information not only to quantify the space-time geometry around it, but to understand physics of processes in plasma in such strong gravitational fields as well. 

The Event Horizon Telescope collaboration (EHT) was the first to scrutinize and confirm experimentally this understanding. Successful imaging of the SMBH in the centre of the galaxy M87$^\ast$ has demonstrated also that firm ultimate conclusion and quantitative characterization urges a much higher angular resolution with interferometric baselines longer than Earth diameter. The maximum angular resolution reached by EHT $\Delta\theta\sim 20~\mu$as at $\lambda=1.3$ mm, i.e. a half of the angular size of the M87$^\ast$ SMBH $\Delta\theta\sim 40~\mu$as being at the edge of the diffraction limit. 

The results obtained with the EHT showed the presence of asymmetry in the shadow of the M87 \citep{2019ApJ...875L...1E, 2019ApJ...875L...5E, EHT2019_p6}. It directly carries the information about the most important parameters of the SMBH space and time metrics, such as the mass $M$, the inclination relative to the observer's line of sight, the spin $a$ \citep{bromley2001, broderick2009, Moscibrodzka2009, dexter2010, Dexter2012, kamruddin2013, Broderick2014}. The scales of such asymmetry are of the order of $\sim5$ $\mu$as. Therefore, going to higher frequencies (above 230~GHz), even this could be challenging, will give a higher angular resolution which is a goal for the next generation ground-base millimeter VLBI array (see, for example, \cite{Doeleman2019}), either going to Space-Earth or Space-Space VLBI program (S-E VLBI or S-S VLBI, correspondingly) is needed to advance the direct experimental examination of General relativity \citep[see discussions in ][]{Roelofs2019,Andrianov2021,Novikov_2021}.  

In our recent paper \cite{Andrianov2021} it was shown that a S-E VLBI experiment with a high elliptical orbit around Earth (HEO) can provide a better image quality of both Sgr~A$^\ast$ and M87$^\ast$ nearby SMBH even in an image dynamical mode, with a higher fidelity and a better nearly proportional to the angular resolution asymmetry test. From this point of view, this is achieved despite of a sparse $(u,v)$ coverage typical for single space based station of S-E VLBI. It seems reasonable to expect that result for experiments with an even higher angular resolution ($\Delta\theta\sim 0.1~\mu$as) at the Sun-Earth Lagrangian libration point L2 (for details see Section \ref{mmtron}). On the other hand though, a common view is that a $(u,v)$ coverage with baselines at L2 and a group of available ground-based stations would be insufficient for imaging with reasonable quality.

Our current paper aims to demonstrate that this is not exactly the case. In fact $(u,v)$ coverage which is very similar to Earth VLBI can be obtained if spacecraft is in the line of sight from the center of ground based array and the target source. A proper signal processing and orbit parameter choice has an ability to provide reasonably convincing conclusions.

\section{Millimetron Observatory} \label{mmtron}
Millimetron observatory will be a 10 meter deployable space telescope that is capable to operate at far infrared, sub-millimeter and millimeter space telescope \citep{MM_2014}. It will have two operating modes: single dish and Space-Earth VLBI. In order to achieve the best required sensitivity for the single-dish, the antenna and onboard instrumentation will be cooled down to 10~K and 4~K correspondingly. 

Scientific tasks for the single-dish mode include the measurement of CMB spectral distortions and magnetic fields, observation of filamentary structure and search for the water trail in the Galaxy.
Regarding the Space-Earth VLBI mode, Millimetron is aimed to study the vicinity of the nearest SMBHs (M87$^\ast$ and Sgr~A$^\ast$) with high angular resolution in a wide frequency range. The sensitivity of the observatory Millimetron will be orders of magnitude better than of Radioastron \cite{Kardashev2013}. The design system temperature of VLBI receivers is shown in Table~\ref{Table_MM}.

\begin{table}
\caption{Parameters of VLBI Mode of Millimetron Space Observatory.}
\centering
\label{Table_MM}
\begin{tabular}{l|c|c}
\hline
Band, (GHz)   	                        & T$_{sys}$, (K)           	& SEFD, (Jy)\\
\hline
33 - 50~GHz                             & < 17~K                    & $\sim$1000\\
84 - 116~GHz                            & < 37~K                    & $\sim$1900\\ 
211 - 275~GHz                           & < 50~K                    & $\sim$4000\\
275 - 373~GHz                           & < 90~K                    & $\sim$5000\\
602 - 720~GHz                           & < 150~K                   & $\sim$10600\\
\hline
\end{tabular}
\end{table}

Millimetron space observatory will be operating in halo orbit around L2 point of the Sun-Earth system. It is a quasi-stable orbit located in the vicinity of L2 point in the plane perpendicular to the ecliptic plane. Such orbital configuration is the only option for single dish observations in terms of thermal and radiation conditions and it will provide the lowest possible temperature of the telescope mirror and thus allow to reach ultimate sensitivity. Therefore, this type of orbit is the main and will be unchanged, and alternative orbit configurations (such as in \cite{Andrianov2021}) could be considered only as additional stages of the mission after the implementation of the main scientific program.

The observatory will have an onboard memory, so that Space-Earth VLBI observations will be conducted without the simultaneous data transfer to the Earth. The signal will be stored in the onboard memory and transferred to the ground after the observation or in the gaps between. The expected volume of this memory is 100 Tb, that corresponds to $\sim$830 minutes of observations with four channels (right and left circular polarizations and two side bands) each 2~GHz bandwidth channel.

\section{Orbit Configuration}

To perform the simulations of S-E VLBI observations we have calculated an orbit around L2 point using the software for navigational and ballistic support developed in Astro Space Center. Orbit calculation method can be divided into the following steps:

\begin{itemize}
     \item Search for analytical halo orbit that meets the necessary requirements for baseline projections, $(u,v)$ coverage and source visibility
     \item Check for stability of the orbit in the implemented force model
     \item Search for orbital regions that provide the best $(u,v)$ coverage
     \item Integrate orbital region using an exact propagator
     \item Calculate $(u,v)$ plane for a given orbital region
\end{itemize}

The force model accounts for such factors as the central gravitational field of the Earth, non-centrality of the Earth gravitational field up to the 8th order (EGM96), as well as the attraction of the solar system bodies: the Moon, the Sun, Mercury, Venus, Mars, Jupiter, Saturn, Uranus, Neptune and Pluto. Additionally, the force model takes into account the contribution of the solar radiation pressure on the spacecraft. The orbit is integrated using the Runge-Kutta method of 7(8)th order \citep{Montenbruck_1992, Schafer_2002}.

\begin{figure*}
    \centering
    \includegraphics[width=\linewidth]{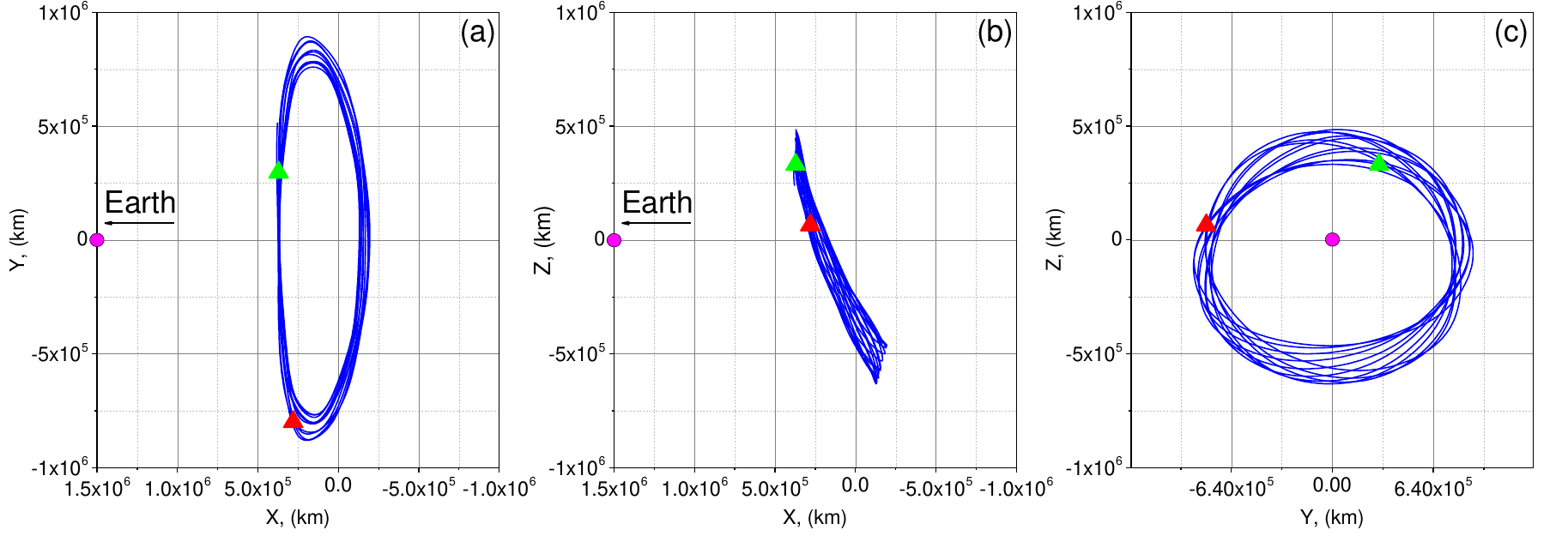}
    \caption{Projection of Millimetron L2 point halo orbit in L2 coordinate system (five orbital periods). (a) -- XY-plane, (b) -- XZ-plane, (c) -- YZ-plane. In the presented coordinate system, the X-axis is directed to the Sun, the Z-axis is perpendicular to the ecliptic plane and directed to the pole of the world, the Y-axis complements. Magenta circle corresponds to the Earth position, red and green triangles correspond to the spacecraft position for the calculated $(u,v)$ coverages for M87$^\ast$ and Sgr~A$^\ast$ correspondingly.}
    \label{fig:fig1}
\end{figure*}

Fig.~\ref{fig:fig1} shows the projections of calculated analytical halo orbit in the XY, XZ and YZ planes. Red and green triangles on Fig.~\ref{fig:fig1} correspond to the regions of orbit (spacecraft position), that provide the $(u,v)$ coverages described below for M87$^\ast$ and Sgr~A$^\ast$ correspondingly. The exit from the ecliptic plane in this orbit does not exceed 400 thousand km. The choice of such an exit from the ecliptic plane increases daily radio visibility of the spacecraft at the Russian tracking stations for regular communication and control sessions with the spacecraft. More details on the orbital configuration can be found in \cite{Rudnitskiy2021}.

\begin{figure*}
    \centering
    \includegraphics[width=0.8\linewidth]{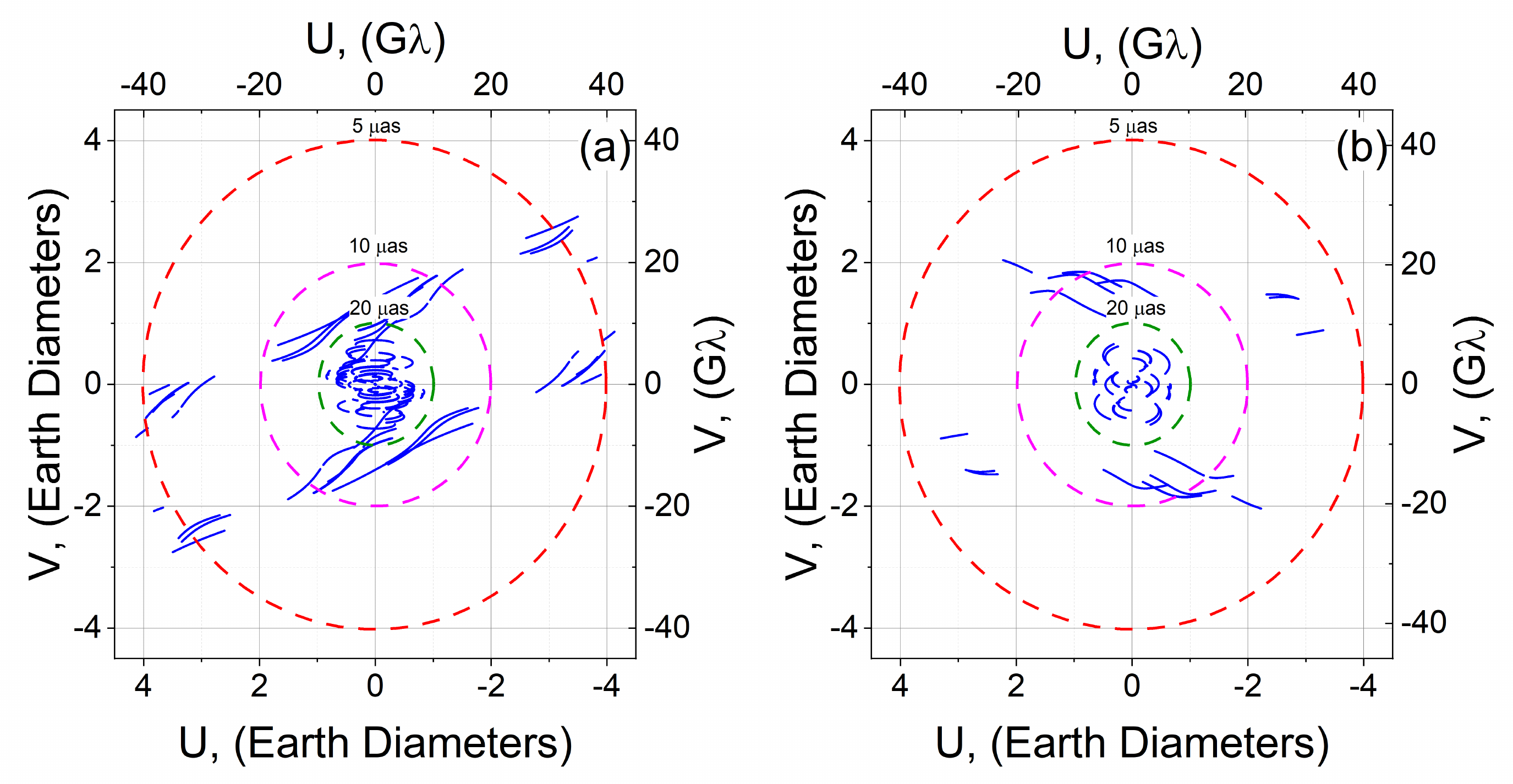}
    \caption{$(u,v)$ coverage for M87$^\ast$ (left) and Sgr~A$^\ast$ (right) for L2 point orbit of Millimetron+EHT configuration. Concentric dashed circles in red, magenta and green correspond to the angular resolutions of 5, 10 and 20 $\mu$as at 230~GHz. Ground telescopes used in calculations are listed in Table~\ref{Table_EHT}.}
    \label{fig:fig2}
\end{figure*}

The goal of the work, as it was mentioned, is to demonstrate the L2 point orbit capabilities of VLBI imaging for two sources (M87$^\ast$ and Sgr~A$^\ast$) using the space-ground interferometer mode of the Millimetron observatory. For this purpose, several regions were selected on the calculated orbit that provide acceptable $(u,v)$ coverage for M87$^\ast$ and Sgr~A$^\ast$ target sources (see Fig.~\ref{fig:fig2}). Coverage was calculated using the EHT ground telescopes (see Table~\ref{Table_EHT}), taking into account the visibility of sources (Sgr~A$^\ast$ and M87$^\ast$) from ground-based and space telescopes. The $(u,v)$ points of space-ground baselines cover the range of baseline projections from 0.8 to 3 Earth diameters for M87$^\ast$ and from 1 to 3 Earth diameters for Sgr~A$^\ast$. This makes it possible to perform the imaging of these sources with the resolution $\sim$3-4 times higher than it was achieved by the EHT. It should be noted that with a further increase in the space-ground baseline projections, the coverage begins to degenerate, the synthesized beam becomes extended along one of the directions, which leads to the degeneration of two-dimensional image to a one-dimensional.

Such orbital regions make it possible to perform VLBI imaging with an adequate $(u,v)$ coverage once a year for each of the selected sources, while the duration of such small baseline projection region in time is several days.

Thus, according to \cite{Rudnitskiy2021}, we demonstrate that the orbit in the vicinity of L2 point of the Sun-Earth, despite the large distance from the Earth (1.5 million km), allows one to obtain small baseline projections of several Earth with, at the same time, relatively good $(u,v)$ coverage for imaging with a resolution that will in any case be better than with ground-based telescopes only. Undoubtedly, the L2 point orbit also makes it possible to obtain extended space-ground baselines, even up more than 100 Earth diameters. In this case, only observations related to the measurement of the visibility function at these spatial frequencies are possible.

As it was previously mentioned, the Millimetron project is designed to operate in two modes: single-dish and VLBI mode. The single-dish mode requires the highest achievable levels of sensitivity. Thus the antenna and the scientific payload of Millimetron observatory will be mechanically cooled down to 10~K and 4~K correspondingly. In this case the orbit around the L2 point of the Sun-Earth system is optimal since the noise contribution from the Earth and the Moon is blocked by the spacecraft heat shields. 

\begin{figure*}
    \centering
    \includegraphics[width=\linewidth]{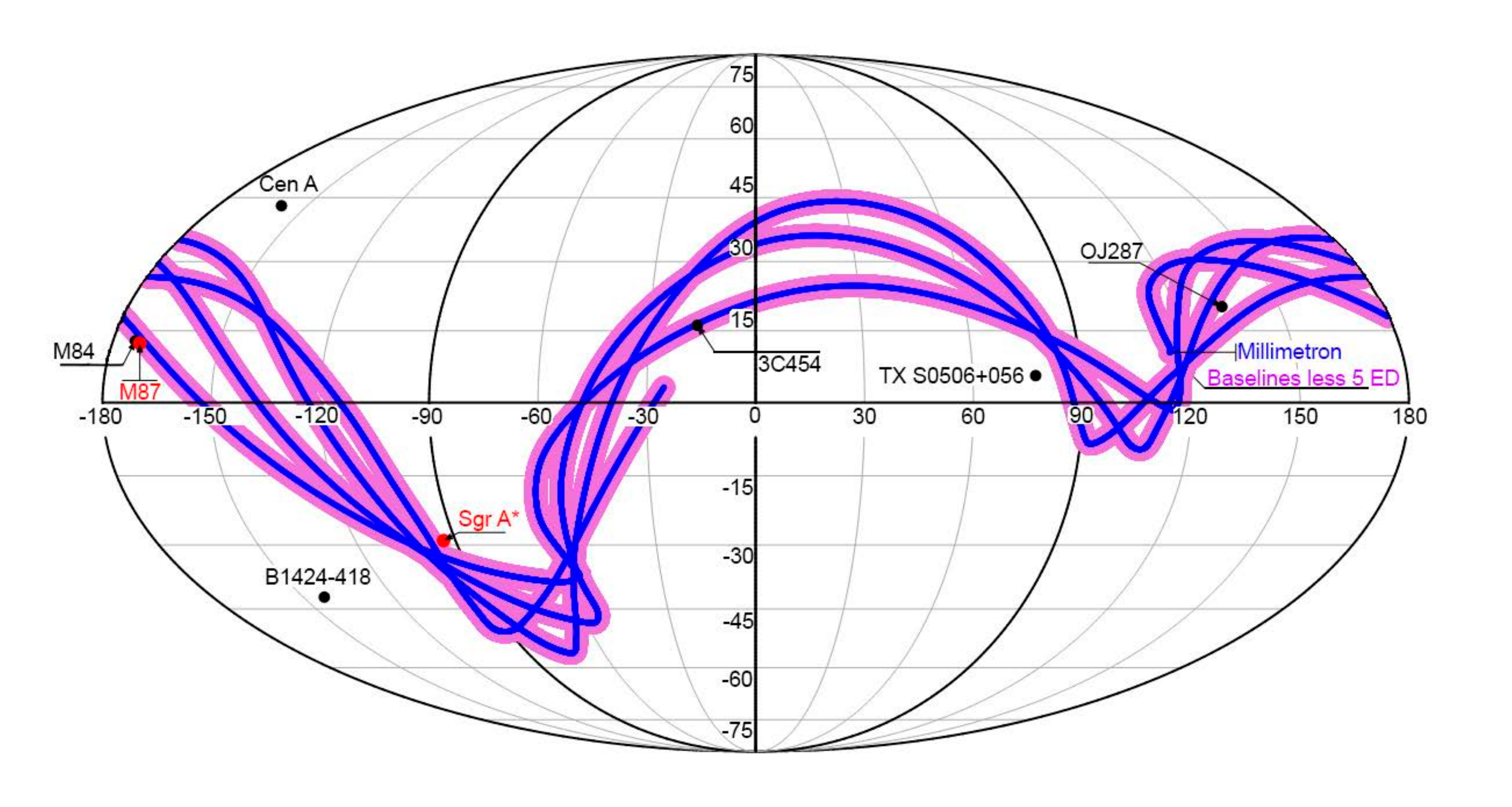}
    \caption{Mollweide projection of Millimetron orbit onto celestial sphere (blue lines) in the GCRF (Geocentric Celestial Reference Frame) for one of the orbits optimized for M87$^\ast$ and Sgr~A$^\ast$ observations. Pink track on picture shows area of baseline projections less 5 Earth diameters.}
    \label{fig:mol}
\end{figure*}

In order to fulfill all the scientific tasks in the VLBI mode it is necessary to have a halo orbit around the libration point of the Sun-Earth system such that would satisfy the necessary requirements for baseline projections and $(u,v)$ coverage for the sources, listed as priority in Millimetron science case.
Thus, the presented orbit can be optimized in the future to reduce the magnitude and number of corrections required to maintain the halo orbit as well as to increase the number of the sources for imaging.
For example, Fig.~\ref{fig:mol} shows a halo orbit that is suitable for S-E VLBI observations with relatively short baseline projections $< 5$ Earth diameters for the sources that are located close to the orbit tracks (blue lines in Fig.~\ref{fig:mol}).

\section{Simulations}
The simulations consisted of the following steps: a setup of general parameters (i.e. bandwidth, observing frequency, integration time, etc.); simulation of the space-ground interferometer geometry with a given orbital configuration and the set of ground telescopes to obtain required $(u,v)$ coverage for M87$^\ast$ and Sgr~A$^\ast$; application of source models (visibility) onto the obtained $(u,v)$ coverage taking into account the telescope sensitivity and phase errors; perform fringe fitting (for the case of phase errors simulations shown in Fig.~\ref{fig:fig8}) and image reconstruction; estimation of image quality.

\subsection{Setup} \label{subsec:setup}

The main goal of presented simulations is to demonstrate the principled capabilities of Millimetron observatory Space-Earth VLBI imaging in L2 point orbit of averaged Sgr~A$^\ast$ and M87$^\ast$. In order to compare the results of L2 orbit imaging with HEO and ground only baseline the parameters of the simulations were taken the same as for \cite{Andrianov2021}, including the configuration of ground array. Namely, bandwidth was set to $\Delta\nu=2$~GHz, total observing time $t=15$ hours, the observing frequency $\nu=230$~GHz and coherent integration time $t_{int}\approx$10~s. 

The simulation took into account the telescope sensitivities and phase errors. A global fringe-fitting procedure was used in order to calibrate these phase errors. More details on the calculation of phase errors is provided below in Section~\ref{subsect:phase_er}. The case of phase errors was simulated separately and only for M87$^\ast$ source as an example. All other simulations accounted for the telescope sensitivity only. Other effects like on-closing or polarimetric leakages were left aside.

\begin{table*}
\caption{Parameters of ground telescopes at 230 GHz \citep{2019ApJ...875L...3E}}
\resizebox{\linewidth}{!}{%
\begin{tabular}{lrrrrr}
\hline
Telescope												&X,~m		&Y,~m			&Z,~m		&SEFD,~Jy		&$D$,~m	\\
\hline
Atacama Large Millimeter Array, Atacama, Chile (ALMA)       					&  2225061.164  	&   -5440057.370    		&   -2481681.150    	&     74 		& 73      	\\
Atacama Pathfinder Experiment, Atacama, Chile (APEX)					&  2225039.530   	&   -5441197.630    		&   -2479303.360    	&   4700 		& 12      	\\
Greenland Telescope, Greenland (GLT)   						&  1500692.000   	&   -1191735.000     		&    6066409.000     	&  5000  		& 12      	\\
IRAM 30-m millimeter radio telescope, Pico Veleta, Spain (PV)					&  5088967.900	&    -301681.600  		&    3825015.800  	&   1900 		& 30      	\\
James Clerk Maxwell Telescope, Hawaii (JCMT)						& -5464584.680   	&   -2493001.170    		&    2150653.980    	&  10500 		& 15      	\\
Large Millimeter Telescope, Mexico (LMT)							&  -768713.964 	&   -5988541.798  	&    2063275.947 	&   4500 		& 50 		\\
Submillimeter Telescope, Arizona, United States (SMT)					& -1828796.200  	&   -5054406.800   		&    3427865.200   	&  17100		& 10      	\\
Submillimeter Array, Hawaii, (SMA)								& -5464523.400  	&   -2493147.080   		&    2150611.750   	&   6200 		& 14.7      	\\
\hline
Kitt Peak National Observatory, Arizona, United States, (KP)$^{2020}$			& -1995678.840  	&   -5037317.697   		&    3357328.025   &  13000 		& 12     	\\
Northern Extended Millimeter Array, Plateau de Bure, France (NOEMA)$^{2020}$ 	&  4523998.400	&     468045.240  		&    4460309.760   &  700  		& 52      	\\
\hline
\end{tabular} 
}
\label{Table_EHT}
\raggedright
$^{2020}$ \small{-- telescopes to be added to the EHT in 2020}.
\end{table*}

All the calculations were performed using the Astro Space Locator Software (ASL) package for VLBI data processing and reduction \citep{ASCCorrelator, ASLCASA, Likhachev2020}.

\subsection{Source Models} \label{sect:models}

Sgr A* simulations use the models and the scattering algorithms the same as in \cite{Andrianov2021}. These are the averaged models calculated and described by \cite{Moscibrodzka2014}. They include a set of models each time-averaged over $\Delta t\approx 3$ hours: \# 16, \# 24, \# 31 and \# 39 in nomenclature of \cite{Moscibrodzka2014}. The models are shown in Fig. \ref{fig:fig6} (leftmost column). The model parameters of Sgr~A$^\ast$ are shown in the Table \ref{Table_model_Sgr}. The difference of the models lies in the inclination angle $i$ between the BH spin and the observer's line of sight, the electron temperature $\Theta_{e,\rm j}=k_BT_e/m_ec^2$ in jet where c is the speed of light, $m_e$ is the mass of the electron, $k_B$ is the Boltzmann constant, $T_p/T_e$ is the ratio of protons to electrons temperatures and $\dot M$ is the accretion rate. The models that were used in simulations has an applied diffractive and refractive scattering with the parameters constrained in \citep{Scattering}. In near infrared the characteristic Sgr~A$^\ast$ demonstrates high time variability with time scale ranges from $0.3$ min to $\sim 245$ min as measured by \cite{Eckart2006, Witzel2018}. Intensity and polarization measurements in THz frequency range show a considerable level ($\sim20\%$ to $\sim 49\%$) of time variability within $\sim 3 - 4$ hours \citep{Mauerhan2005, Marrone2006, Marrone2008}. So our time-averaged models represent an idealized scenario. We assumed that after 24 hours of integration, we will get a certain average image that will be in a good correspondence with averaged GRMHD model. In \cite{Andrianov2021} it was demonstrated that static image reconstruction of averaged model and dynamic image reconstruction when the underlying data is evolving in time are in good agreement with each other, so in this paper we will use this assumption.

For M87$^\ast$ source the model that is used in simulations is described by \cite{johnson2019universal}. This model corresponds to magnetically arrested disk (MAD) with the following parameters: black hole mass $M=6.2\times 10^9M_\odot$, spin $a=0.94$ and the angle between the observer and the direction of the jet $\theta_{obs}=163^\circ$. The mass accretion rate matching with observable flux at frequency 230GHz. It is a time-averaged
model with parameters chosen to be consistent with the EHT data of 2017 (\cite{2019ApJ...875L...5E}) as specified in \citep{johnson2019universal}. The time variability of the central region in M87$^\ast$ is months in the radio range \cite{Nagakura2010}. So to obtain time-averaged image of the GRMHD models, Johnson performed over 100 snapshots during time 1 year. The time-averaged model is shown in Fig. \ref{fig:fig4} (a). The motivation to use Johnson models is that this model qualitatively describes observations in M87$^\ast$, see \cite{2019ApJ...875L...1E,2019ApJ...875L...2E,2019ApJ...875L...3E,2019ApJ...875L...4E,2019ApJ...875L...5E}.

But there are other models that are consistent with the observations. Therefore, in addition to Johnson models, we have considered models with other physical parameters. These models can also describe observational manifestations of black hole shadows in M87$^\ast$.
To model the GRMHD model, the free code harm2d \citep{Gammie2003,Noble2006} with a dimension of 2.5 D was used. The axis-symmetric torus-like distribution of an ideal electron-proton plasma around a rotating black hole was considered. The initial state was set as a Fishbone-Moncrief \citep{Fishbone1976} disk with the following parameters: the adiabatic index $\Gamma=4/3$, the black hole rotation parameter $a=0.95$, the inner radius $r_{in}=6$, the radius of maximum pressure $r_{max}=12$ and the parameter beta $\beta=100$, which is the ratio of gas pressure to magnetic pressure. This disk corresponds to a disk with standard and normal evolution (SANE), more details see \cite{Chernov}. This is the main difference between these models and the Johnson's model, which corresponds to the MAD type.

The free code "grtrans"\space\citep{Dexter2016} is used to construct images of black holes. It is supposed that the observer is located at an angle $i = 30^\circ$ to the axis of rotation of the black hole. It is assumed that relativistic electrons emit synchrotron radiation, taking into account absorption and Faraday effects (rotation and conversion). The distribution function of the emitting electrons
was chosen to be equal to the relativistic thermal (Maxwell) distribution function
with the electron temperature determined through the proton temperature. Three models (\#1, \#2 and \#3 correspondingly) with the electron temperature equal to $T_e=T_p/3$, $T_e=T_p$, and $T_e=T_p/2$ are considered, where $T_p$ - is the temperature of the proton. The mass of the black hole is given as $M=6.5\times 10^{9}M_\odot$, where $M_\odot$ is the mass of the Sun and the distance $D\approx16.8$ Mpc \citep{M87PaperVI}. For each model, the radiation flux is normalized at a frequency of $\nu=230$ GHz by an amount equal to $J_{230}=1$, $J_{230}=0.7$ and $J_{230}=0.8$ Jy accordingly. Images of black holes are constructed at a frequency of $\nu=230$ GHz with a resolution of 2500~$\times$~2500 points and the size of field of view is 26 $\times$ 26 $GM/c^2$. Each image has a distinct photon ring. A more detailed description of the models can be found in \cite{Chernov}.

\subsection{Sensitivity and Fringe Detectability} \label{subsect:SEFD}
Data on the telescope sensitivity were then superimposed on the simulated visibility amplitude from Section~\ref{sect:models}. The sensitivity of the ground telescopes used is presented in the Table~\ref{Table_EHT} and is expressed by the $SEFD$ parameter in Jy (see \cite{2019ApJ...875L...2E}). Millimetron sensitivity for 230~GHz is estimated as $SEFD_{MM}=4000$ Jy.

Thus for the most sensitive space-ground baseline Millimetron-ALMA the confident 5$\sigma$ flux detection limit for $\Delta\nu=2$ GHz and $\tau=$10 sec can be estimated as $\approx$ 20 mJy. In the further discussion the sensitivity of this baseline can be considered as an informal cutoff line for the sensitivity.

To estimate the influence of the possible fringe non-detection which may be related to the loss of sensitivity or telescope failures, we have considered a separate case of simulation with M87 source model. In these simulation a set of $(u,v)$ coverages was examined. Several cases were considered when not all telescopes performed successful observation or fringe detection leading to the loss of points in the $(u,v)$ coverage.

\subsection{Phase Errors} \label{subsect:phase_er}
As is known, observations in the millimeter range using ground-based VLBI methods are limited by the influence of the atmosphere which introduces significant phase distortions with increasing frequency.

In order to evaluate the influence of phase errors on the quality of image reconstruction, phase noise was applied to the simulated data of the M87$^\ast$ source. The phase error $\Delta \phi = \Delta \phi_{0} + \Delta \phi_{rms}(t)$ was added to the phase of the model visibility function. Here, $\Delta \phi_{0}$ is a random number from 0 to $360^\circ$, which corresponds to the uncertainty in the absolute initial phase value of a given antenna, $\Delta \phi_{rms}(t)$ reflects the effect of the atmosphere phase noise on the phase of the visibility function. Phase itself changes in time, while the total phase RMS does not change in time. The phase noise of the atmosphere is modeled under the assumption of Gaussian random fluctuations with a root-mean-square deviation \citep{Thompson2017, 2019ApJ...875L...3E, Holdaway1999}. The RMS phase values for the EHT antennas are shown in Table \ref{Table_EHT_Ph}. The estimates were based on the average atmospheric characteristics at the EHT telescope sites \citep[][]{2019ApJ...875L...3E, 2019ApJ...875L...4E}. 

\begin{table}
\centering
\caption{Estimated phase noise for the EHT antennas at 230 GHz.}
\label{Table_EHT_Ph}
\begin{tabular}{lr}
\hline
Telescope												&$\Delta \phi_{rms}$,~deg/sec	\\
\hline
ALMA, Atacama, Chile       					            &  1.1517  \\
APEX, Atacama, Chile					                &  1.1517  \\
GLT, 12 m, Greenland                                    &  1.4396 \\
PV (IRAM 30 m), Pico Veleta, Spain				    	&  2.8793  \\
JCMT, Hawaii,USA                                        &  3.8390  \\
LMT 50 m, Mexico							            &  2.3034  \\
SMT, Arizona, USA					                    &  3.8390  \\
SMA, Hawaii, USA								        &  2.3034  \\
\hline
Kit Peak 10m, Arizona, USA      			            &  2.8793  \\
NOEMA, Plateau de Bure, France 	                        &  1.6453  \\
\hline
\end{tabular} 
\end{table}

\begin{table}
\caption{Parameters of Sgr~A$^\ast$ models \citep{Moscibrodzka2014}.}
\centering
\label{Table_model_Sgr}
\begin{tabular}{l|c|c|c|c}
\hline
models  & i  & $\Theta_{e,jet}$ & $T_p/T_e$ disk & $\dot{M}$ [$M_{\odot} yr^{-1}$ ]    	\\
\hline
\# 16   & $60^\circ$ & 10  &  5 & 3.9$\times 10^{-9}$          \\
\hline
\# 24   & $60^\circ$ & 20  & 20 & 4.2$\times 10^{-8}$           \\ 
\hline
\# 31   & $30^\circ$ & 10  &  5 & 5.6$\times 10^{-9}$         \\
\hline
\# 39   & $30^\circ$ & 20  & 20 & 4.1$\times 10^{-8}$        \\
\hline
\end{tabular}
\end{table}

According to \cite{2019ApJ...875L...3E,2019ApJ...875L...4E}, there are certainly also systematic amplitude gain errors at EHT ground telescopes, but in this simulations we did not take them into account. In the same work (\cite{2019ApJ...875L...4E}) there is a description of the methods that operate directly on robust closure phases, including RML methods, that do not require phase self-calibration. In further research, it is also planned to use these methods.

\subsection{Visibility Calculation}
The described models then were applied as the distribution of complex visibilities taking into account the sensitivity of telescopes and phase noise (for the case of phase simulations). To calculate the visibility function each pixel of the model image was considered as a delta function that can be characterized by the parameter of amplitude magnitude and $(x, y)$ displacement with respect to the center of the image. The complex visibility that correspond to a certain point on the $(u,v)$ plane was calculated as the sum of the Fourier images of delta functions that form the initial model image. Amplitude errors were simulated by adding the thermal Gaussian noise to the visibility amplitude. The amplitude of this thermal noise was set proportional to the sensitivity (SEFD) of the pair of telescopes that form corresponding baseline. Phase errors were simulated by adding to the visibility phase the model shear visibility function characterizing the state of the atmosphere at each of the pair of telescopes that form the corresponding baseline (see Section \ref{subsect:phase_er}).

\subsection{Image Reconstruction}

In presented simulations we use two methods of image deconvolution of the original models: CLEAN and maximum entropy method (MEM) and the image deconvolution was performed with Astro Space Locator software package \citep{Likhachev2020}. Each of these methods has its own advantages and disadvantages, described in sufficient detail in the literature (see, for example, \cite{Thompson2017}). Since the black hole shadow model is, generally speaking, an extended source, MEM is the most appropriate deconvolution method. On the other hand, the model can contain small enough elements (for example, flares, bright compact spots, etc.) and for deconvolution of such image details CLEAN is a rather effective method.

Note that in comparison with linear reconstruction algorithms (CLEAN), the solutions obtained on the basis of the MEM (see \cite{Addario1973,Wernecke1977}) provides higher quality restored images for extended and sources with complex structure. On the other hand, MEM requires a solution of non-linear equations complex system. The target entropy functional is essentially nonlinear and can be used to extrapolate the values of the visibility function outside the bounded $(u,v)$–plane (the so-called "super-resolution"). 

In addition, as will be shown below, the use of both CLEAN and MEM deconvolution methods provide a reliable estimate of the quality of the reconstructed image (fidelity).

We used the following parameters for CLEAN image reconstruction: a uniform weighing, uniform grid kernel function, no tapering and CLEAN was performed with non-negative components.

For the MEM image reconstruction method the regularization function $J$ \citep{1986ARA&A..24..127N}:

\begin{equation}\label{eq:mem} 
J = - \alpha_1 \sum_{i=1}^{n^2}I_{i}^{'} \log I_{i}^{'} - \alpha_2 (\frac{1}{2N}\sum_{j=1}^{N}\frac{(I_{Bj} - I_{Bj}^{'})^2}{\sigma_{Bj}^{2}} - 1),
\end{equation}

have been maximized. Here $n^2$ is the number of pixels on the restored image $I^{'}$, $N$ is the number of measured visibilities $I_B$, $I_B^{'}$ is the visibilities obtained from the Fourier transform of the restored image $I^{'}$, $\sigma_{Bj}$ is the noise estimate on the visibilities $I_Bj$. The first term in the function $J$ is responsible for the entropy function, and the second for the goodness-of-fit test statistic that compares the visibilities of the restored image $I^{'}$ to the measured data. The mixing coefficient $\alpha_1$ and $\alpha_2$ controls the weighting between the entropy term and the data term. For the MEM images in this paper $\alpha_1 = 1, \alpha_2 = 50$.

As a priori image for the MEM method a Gaussian with a half-width $\sigma = w/(2\sqrt{2} ln 2)$ and the total M87$^\ast$ flux obtained from the CLEAN method was used. Here $w$ is the size of field of view. 

The field of view for the images were the following: 100 $\times$ 100 $\mu$as for M87$^\ast$ and 200 $\times$ 200 $\mu$as for Sgr~A$^\ast$. For CLEAN method the map size was 1024 $\times$ 1024 pixels with pixel size of $\sim$0.1 $\mu$as for M87$^\ast$ and $\sim$0.2 $\mu$as for Sgr~A$^\ast$. For MEM the map size was 512 $\times$ 512 pixel with pixel size of $\sim$0.2 $\mu$as.

\subsection{Image Quality}
Image quality was estimated using  image fidelity, SSIM and sharpness methods as presented in \cite{Andrianov2021}. The fidelity quality criterion (or fidelity measure $F$) is described as follows:
\begin{equation}\label{eq:fidelity} 
F = \frac{MAX(M_i)}{\sqrt{\frac{1}{n}\sum_{i=1}^{n}(I_i - M_i)^2}},
\end{equation}
where $F$ is fidelity, $M_i$ is the intensity at the $i$-th pixel in the model image, $I_i$ is the intensity at the corresponding $i$-th pixel in the reconstructed image, $n$ is a number of pixels in the image.

The measure of structural similarity (SSIM) compares local patterns of pixel intensities that have been normalized for luminance and contrast. Consider $A_{1...n},B_{1...n}$ are the arrays of pixel intensity values of two images: model ($A$) and analyzed image ($B$). Both images are arranged in a linear array such as array index correspond to the same position in the image for both $A$ and $B$. Both arrays must have the same length $n$. Then, the mean value $\mu$ of pixel array (image) will be:

\begin{equation}\label{eq:mean} 
\mu = \frac{1}{n}\sum^{n}_{k=1}A_{k},
\end{equation}
and the standard deviation of array $A$:
\begin{equation}\label{eq:rms} 
\sigma = \sqrt{\frac{1}{n}\sum^{n}_{k=1}(A_{k}-\mu_{A})^{2}}.
\end{equation}
Then the SSIM will be calculated as:
\begin{equation}\label{eq:ssim}
SSIM_{A,B} = \frac{2\mu_{A}\mu_{B}}{\mu_{A}^2+\mu_{B}^2}\frac{2\sigma_{A}\sigma_{B}}{\sigma_{A}^2+\sigma_{B}^2}\frac{\sigma_{A,B}}{\sigma_{A}\sigma_{B}},
\end{equation}
where $\sigma_{A,B}$ -- the sharpness of the inner edge defined as the standard deviation of the difference between the reconstructed image and the model in the cross section along the right ascension axis:
\begin{equation}\label{eq:sharpness}
\sigma_{A,B}=\frac{1}{n}\sum^{n}_{k=1}(A_{k} - \mu_{A})(B_{k} - \mu_{B}).
\end{equation}

Note, that SSIM measure used here as human independent measure of this parameter. The SSIM index value ranges from 0 to 1, where index value is equal to 1 for two identical images and equal to 0 for two different images. It should be noted that SSIM measure corresponds well with human perception of the conform of two images \citep{Wang2004}. The model image was taken for SSIM as a reference image.

\begin{figure*}
    \centering
    \includegraphics[width=\linewidth]{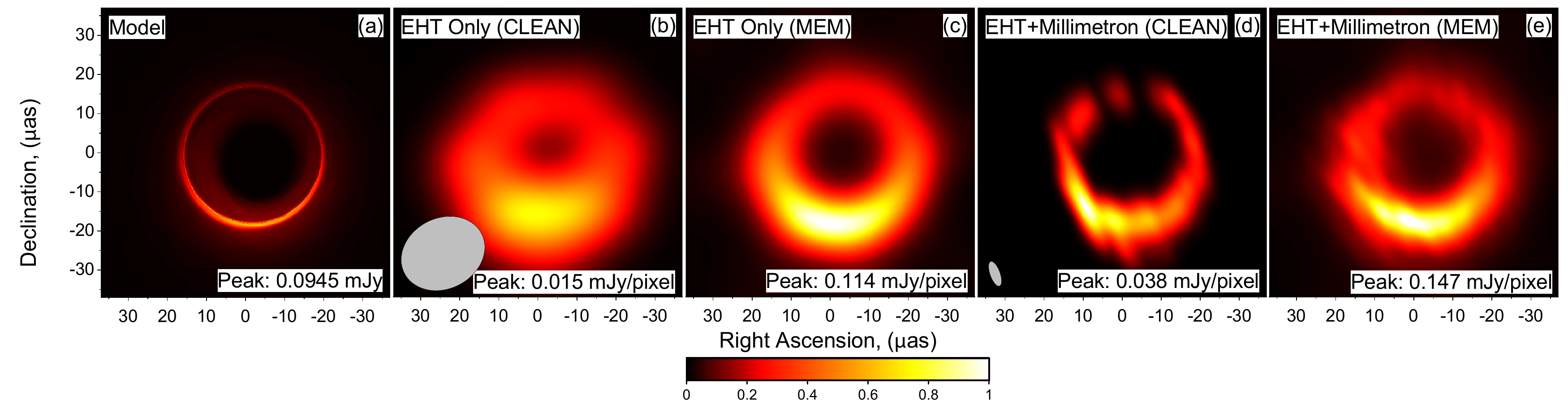}
    \caption{Simulated observations of M87$^{\ast}$ for L2 point orbit:(a) -- M87$^\ast$ model by Johnson et al. This model corresponds to magnetically arrested disk (MAD) with the following parameters: black hole mass $M=6.2\times 10^9M_{\odot}$, spin $a=0.94$ and the angle between the observer and the direction of the jet $\theta_{obs}=163^{\circ}$. The mass accretion rate match the observable flux at frequency 230~GHz. It is a time-averaged model (over 100 snapshots during 1 year) with parameters chosen to be consistent with the EHT data of 2017; (b) and (c) -- EHT only images obtained using CLEAN and MEM methods correspondingly, (d) and (e) -- images obtained by Millimetron in L2 point orbit together with EHT using CLEAN and MEM methods correspondingly. Estimated beam size for EHT only image is 18 $\times$ 16 $\mu$as and for EHT+Millimetron is 8.0 $\times$ 3.5 $\mu$as. No amplitude or phase systematic errors were applied.}
    \label{fig:fig4}
\end{figure*}

\begin{figure*}
    \centering
    \includegraphics[width=\linewidth]{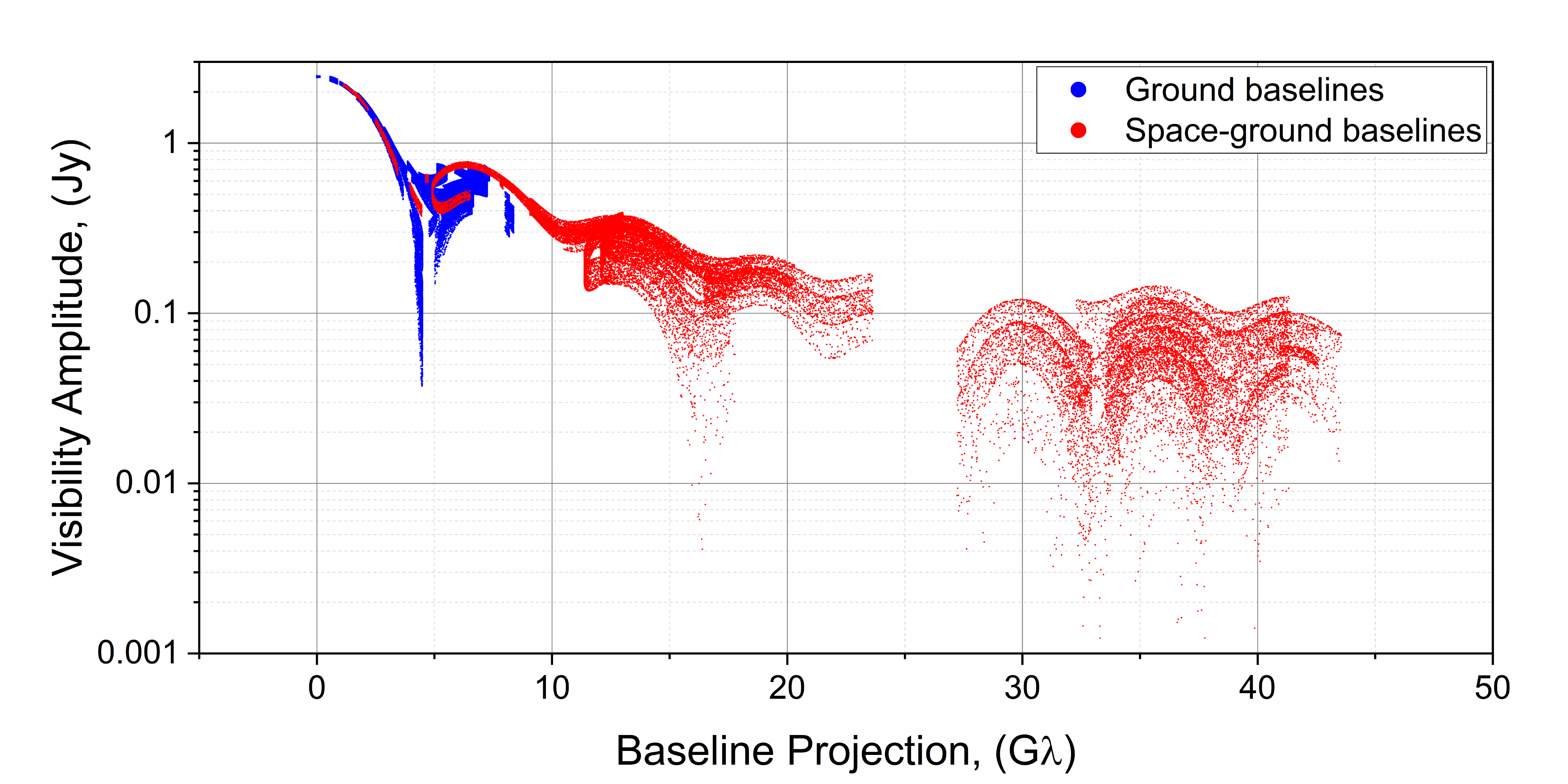}
    \caption{Visibility vs. baseline projection for simulated observations of M87$^\ast$ (model by Johnson et al.). Each point corresponds to 10~s integration time. Points include thermal noise error.}
    \label{fig:vis_m87}
\end{figure*}

\begin{figure*}
    \centering
    \includegraphics[width=\linewidth]{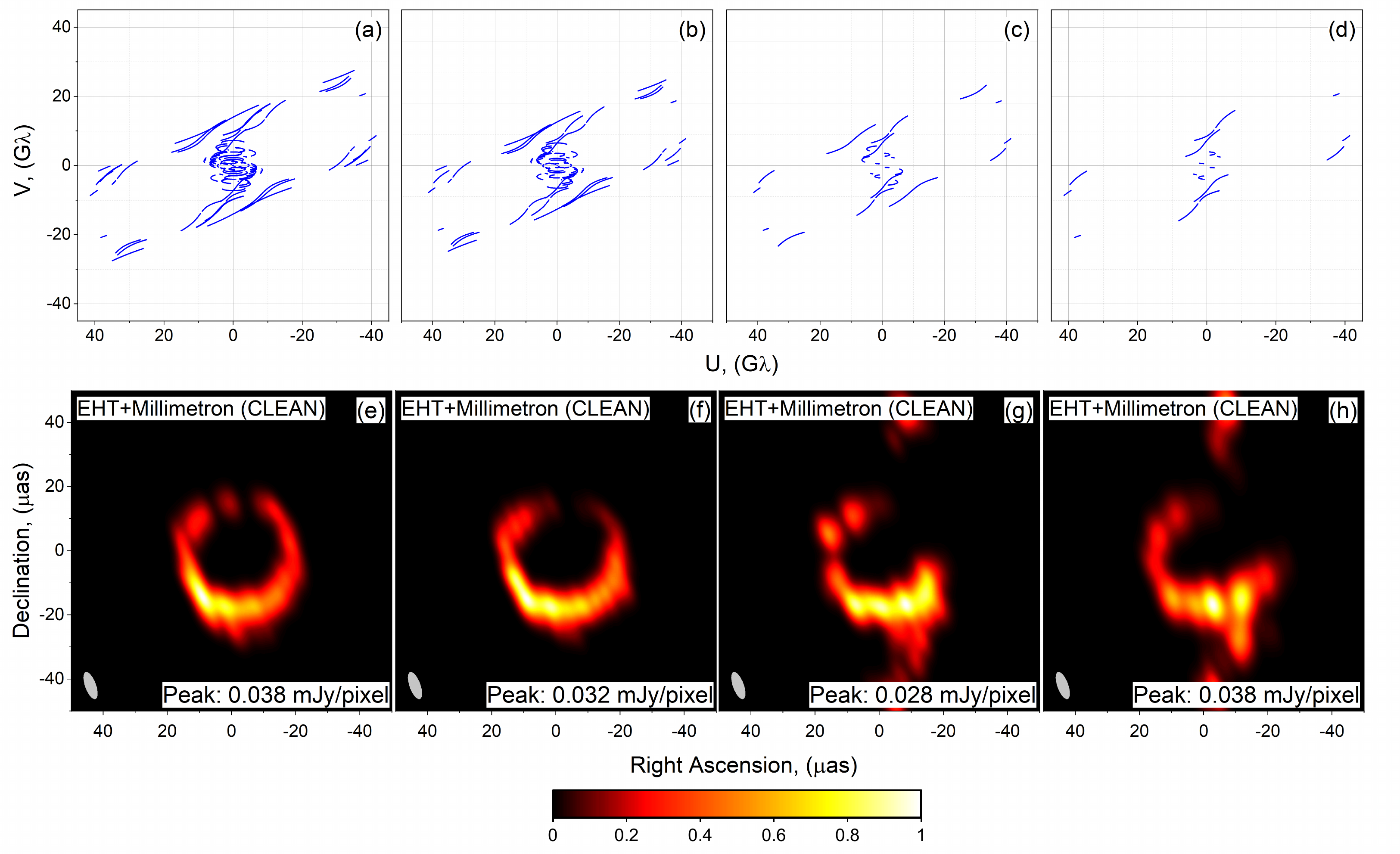}
    \caption{Example of resulting images (a)-(d) of M87$^\ast$ (model by Johnson et al.) depending on the quality of corresponding $(u,v)$ coverage (e)-(h). The following configurations of ground telescopes have $(u,v)$ coverages: (a) -- ALMA, APEX, GLT, JCMT, Kitt Peak, LMT, NOEMA, PV, SMA, SMT; (b) -- ALMA, APEX, GLT, LMT, NOEMA, PV, SMA; (c) -- ALMA, LMT, NOEMA; (d) -- ALMA, LMT. Estimated beam size is: (e) -- 8.08 $\times$ 3.52 $\mu$as , (f) -- 8.11 $\times$ 3.52 $\mu$as, (g) -- 8.6 $\times$ 3.68 $\mu$as, (h) -- 10.2 $\times$ 3.8 $\mu$as. No amplitude or phase systematic errors were applied.}
    \label{fig:fringe_m87}
\end{figure*}

\begin{figure*}
    \centering
    \includegraphics[width=0.9\linewidth]{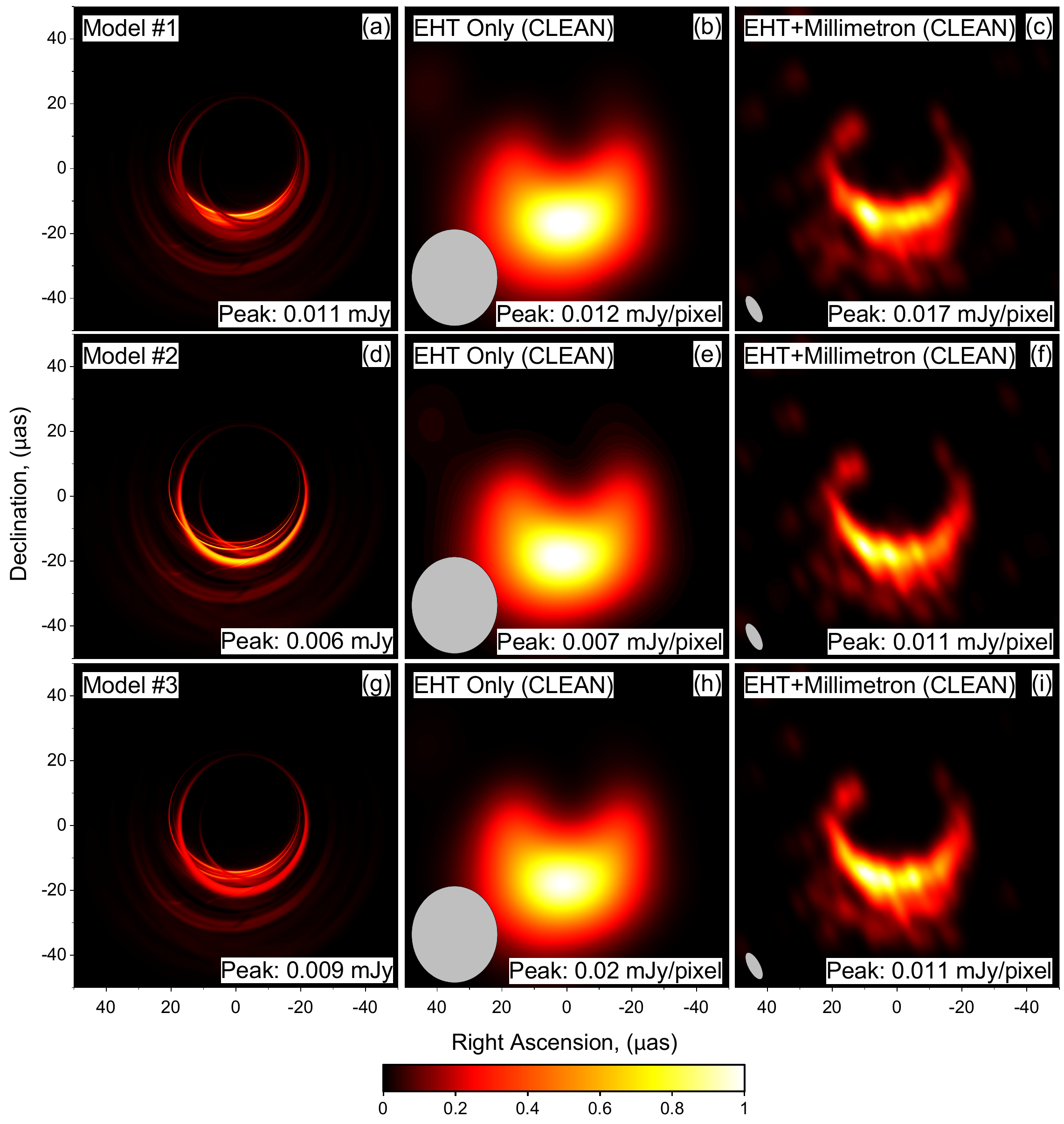}
    \caption{Simulated observations of M87$^\ast$ for L2 point orbit (SANE class models by Chernov et al.). Left column -- models, middle column --  EHT only images, right column -- images obtained by Millimetron in L2 point orbit together with EHT. Estimated beam size for EHT only image is 22 $\times$ 17 $\mu$as and for EHT+Millimetron is 8.08 $\times$ 3.52 $\mu$as. No amplitude or phase systematic errors were applied.}
    \label{fig:fig5}
\end{figure*}

\begin{figure*}
    \centering
    \includegraphics[width=0.9\linewidth]{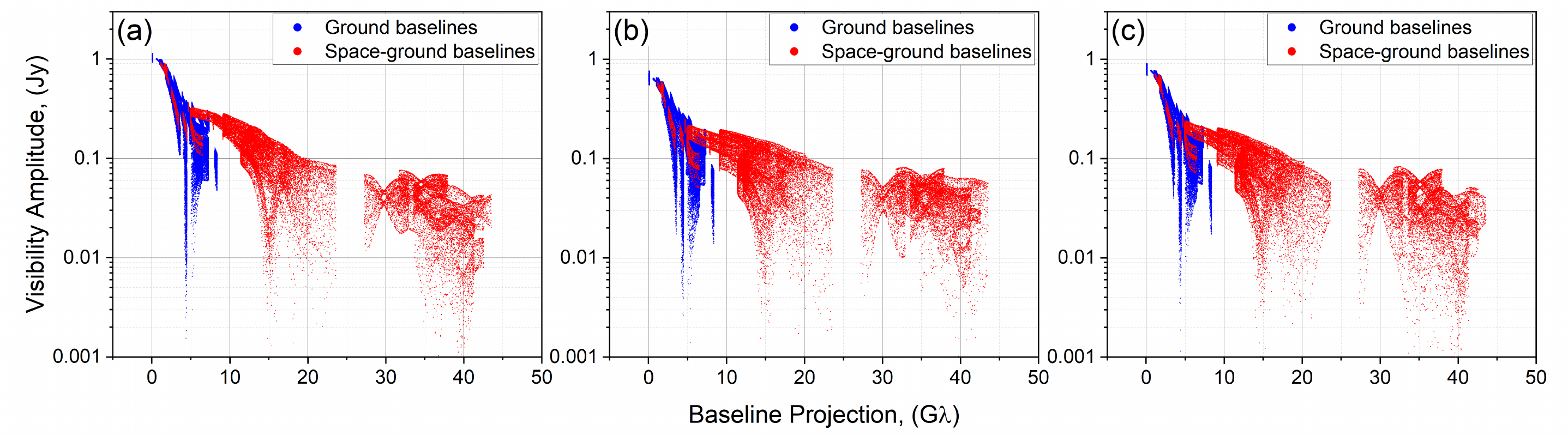}
    \caption{Visibility vs. baseline projection for simulated observations of M87$^\ast$ (SANE class models by Chernov et al.): (a) -- model \#1, (b) -- model \#2, (c) -- model \#3. Each point corresponds to 10 s. integration time. Points include thermal noise error.}
    \label{fig:fig5b}
\end{figure*}

\begin{figure*}
    \centering
    \includegraphics[width=0.8\linewidth]{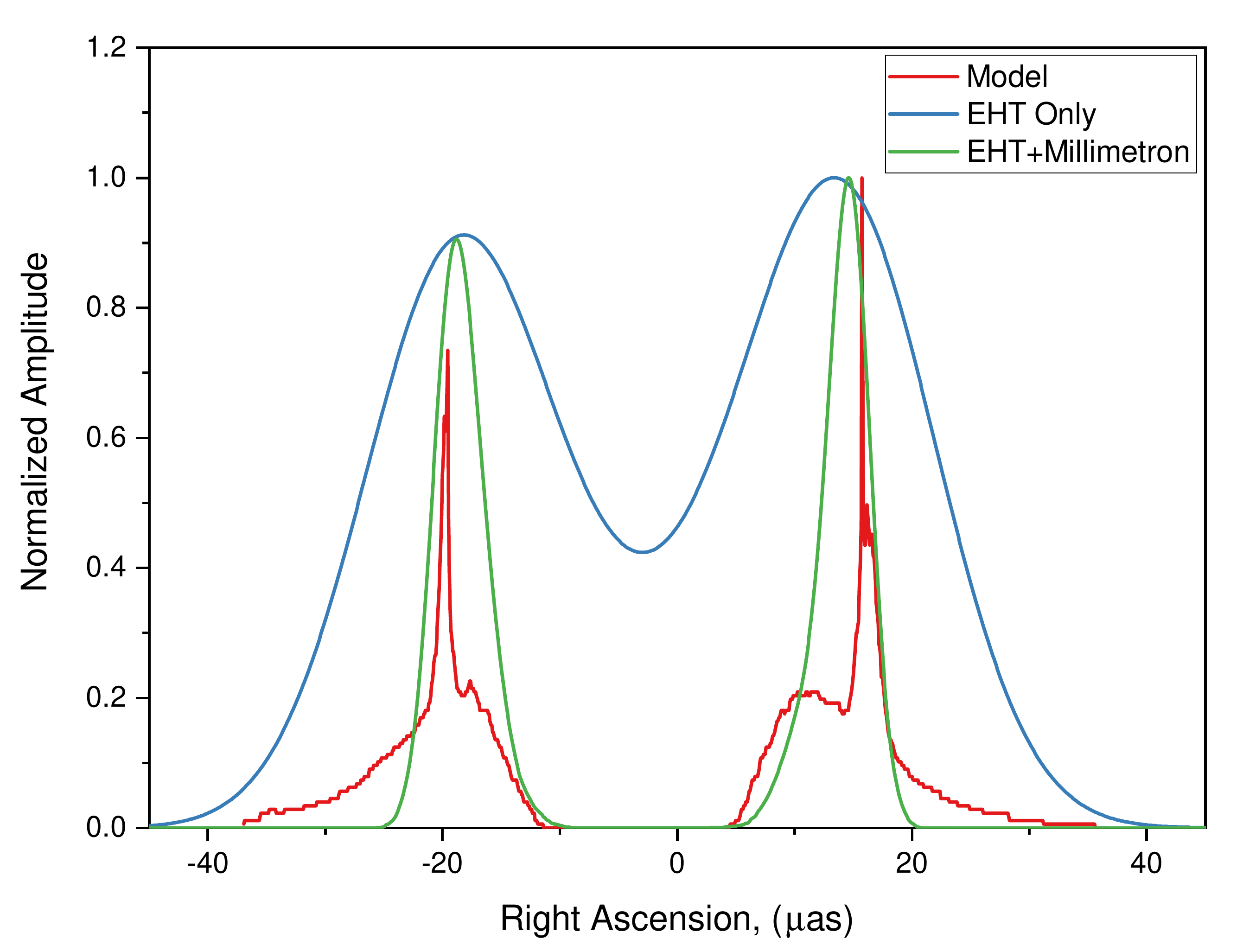}
    \caption{One-dimensional profiles for CLEAN images of M87 (Fig.~\ref{fig:fig4} (a), (b) and (d)) along the right ascension axis at zero declination value. Red line corresponds to the profile of the model, blue line corresponds to the profile of EHT only image and green line corresponds to the profile of EHT+Millimetron image. Amplitudes of each profile were normalized to its peak values.}
    \label{fig:fig5c}
\end{figure*}

\begin{figure*}
    \centering
    \includegraphics[width=0.9\linewidth]{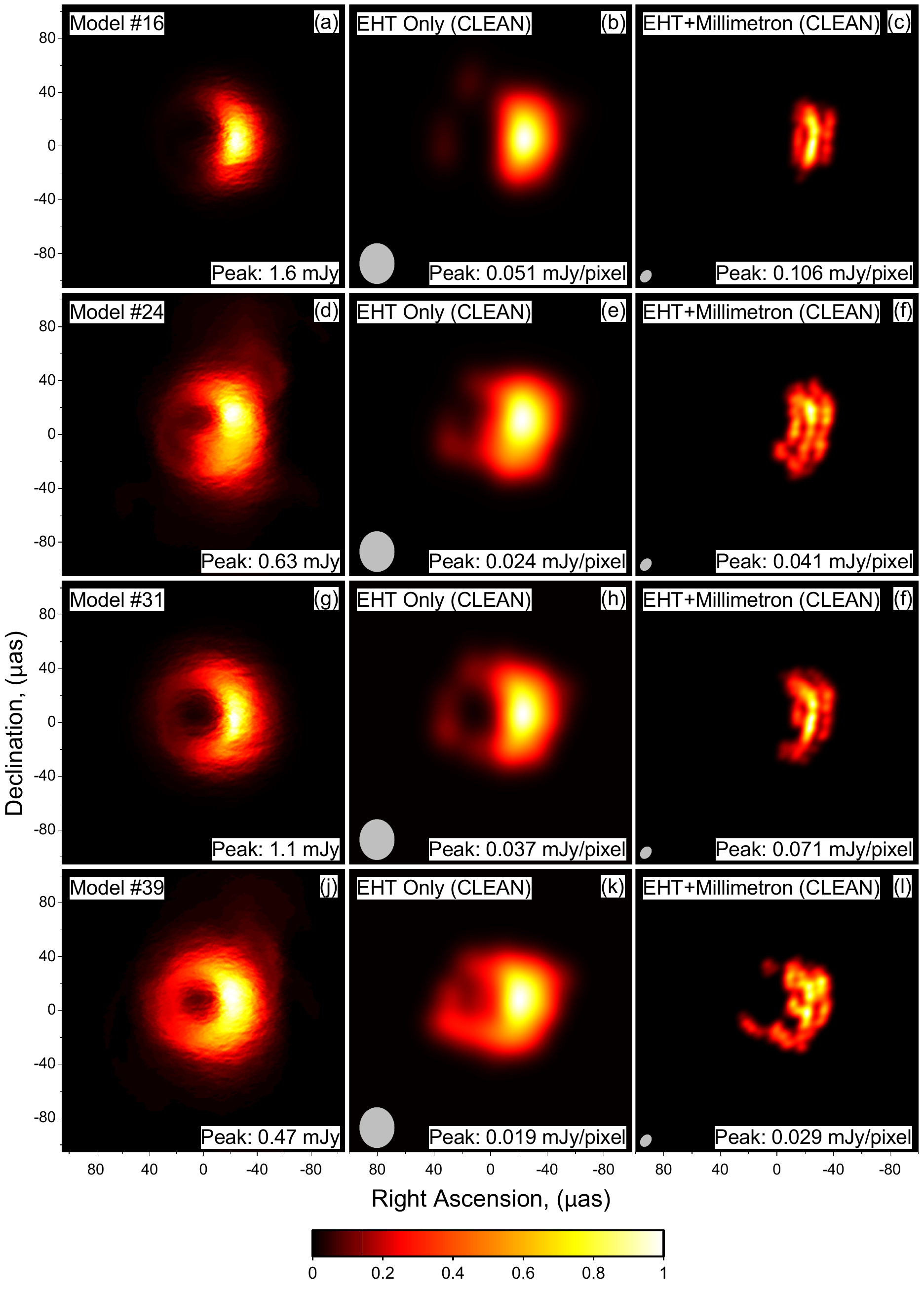}
    \caption{Simulated observations of Sgr~A$^\ast$ (models by Moscibrodzka et al.) for L2 point orbit. Left column -- models, middle column -- EHT only images, right column -- images obtained by Millimetron in L2 point orbit together with EHT. The models were averaged in time over 3 hours with a single frame duration of $\approx$221 s.} Synthesized beam for EHT only image is 21 $\times$ 19 $\mu$as and for EHT+Millimetron 7.3 $\times$ 5.3 $\mu$as correspondingly. \textbf{No amplitude or phase systematic errors were applied.}
    \label{fig:fig6}
\end{figure*}

\begin{figure*}
    \centering
    \includegraphics[width=0.95\linewidth]{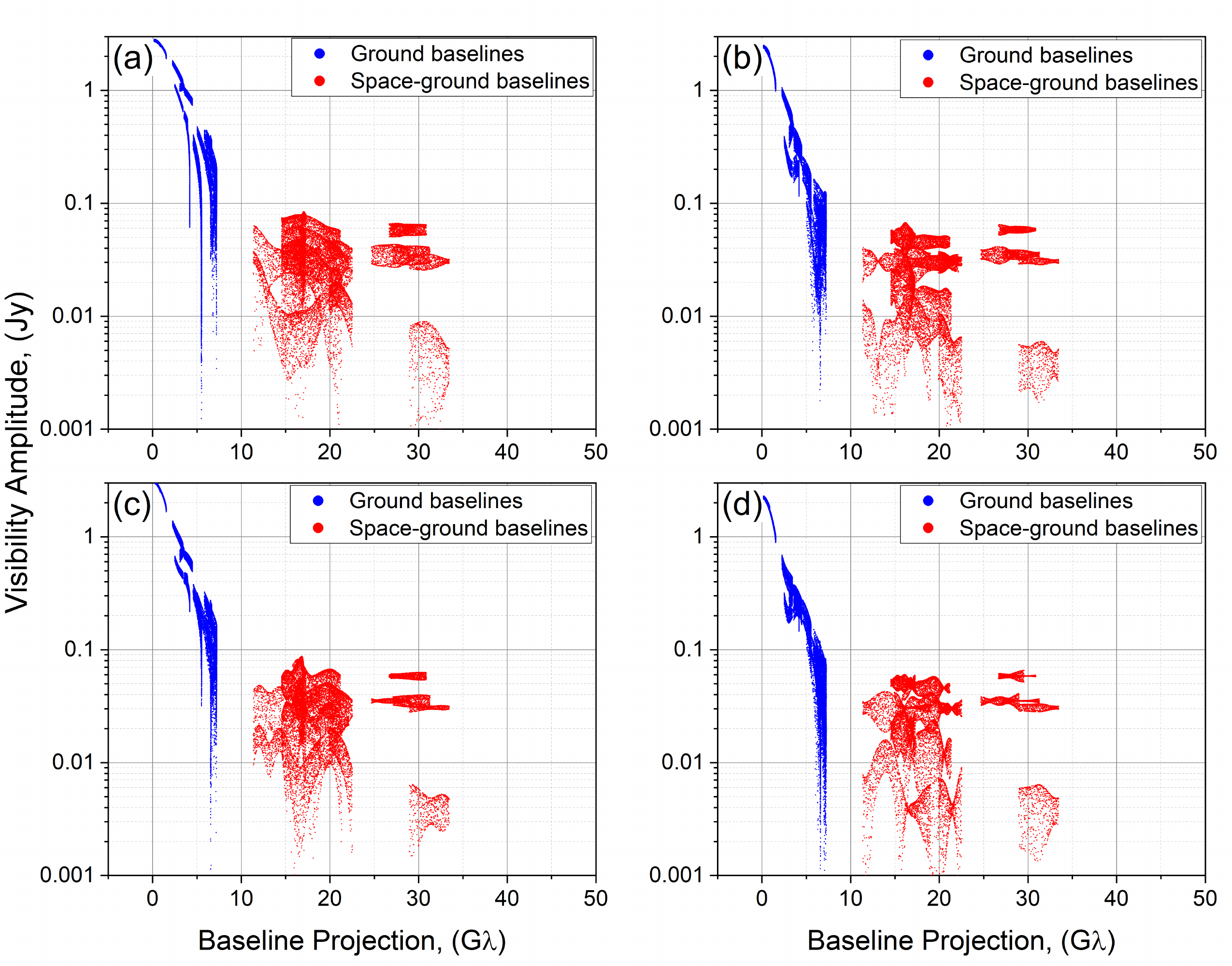}
    \caption{Visibility vs. baseline projection for simulated observations of Sgr~A$^\ast$: (a) -- model \#16, (b) -- model \#24, (c) -- model \#31, (d) -- model \#39. Each point corresponds to 10 s. integration time. Points include thermal noise error.}
    \label{fig:vis_sgr}
\end{figure*}

\begin{figure*}
    \centering
    \includegraphics[width=\linewidth]{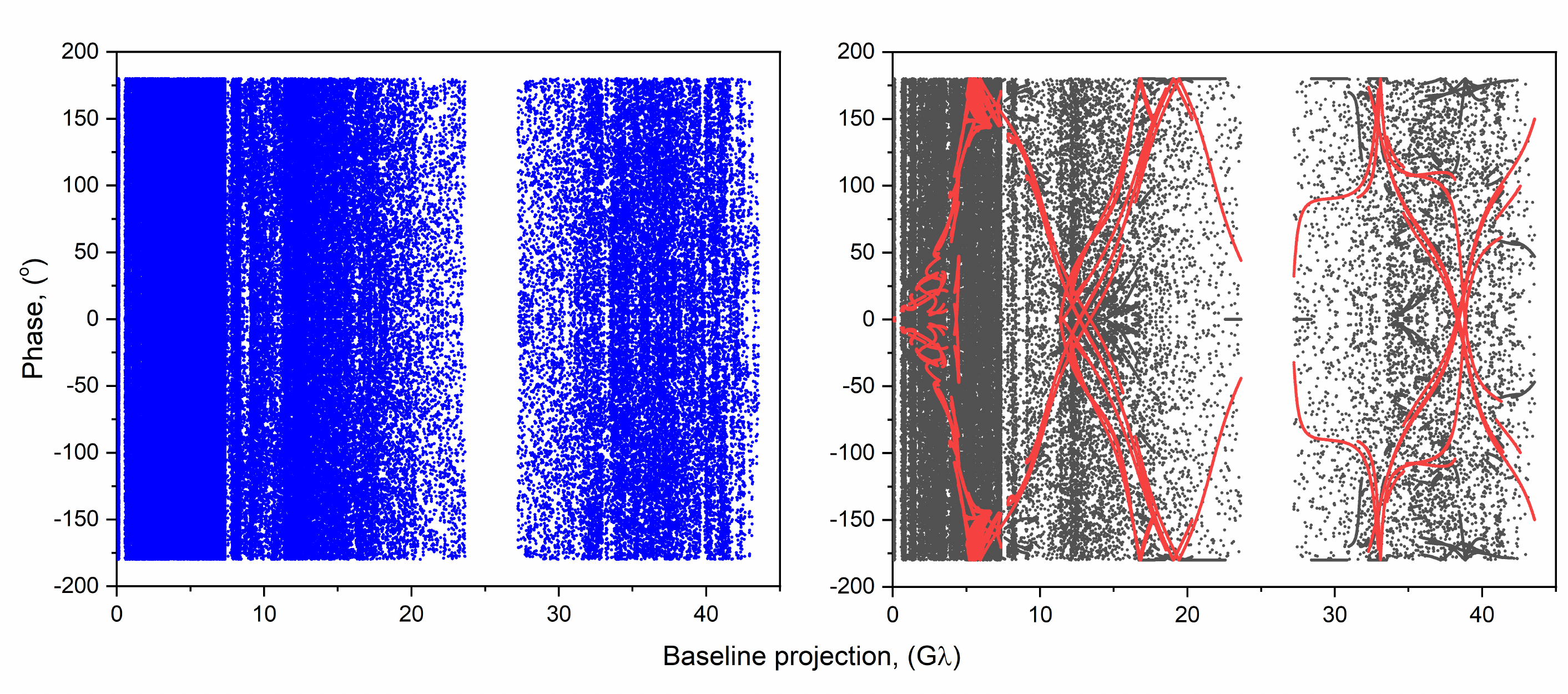}
    \caption{Phase vs. baseline projection for simulated observations of M87$^\ast$. Left panel shows introduced phase noise into data. Right panel corresponds to the resulting phase obtained from noisy data using fringe fitting procedure with M87$^\ast$ model. Red dots on the right panel correspond to initial simulated phase without introducing the noise.}.
    \label{fig:fig7}
\end{figure*}

\begin{figure*}
    \centering
    \includegraphics[width=0.8\linewidth]{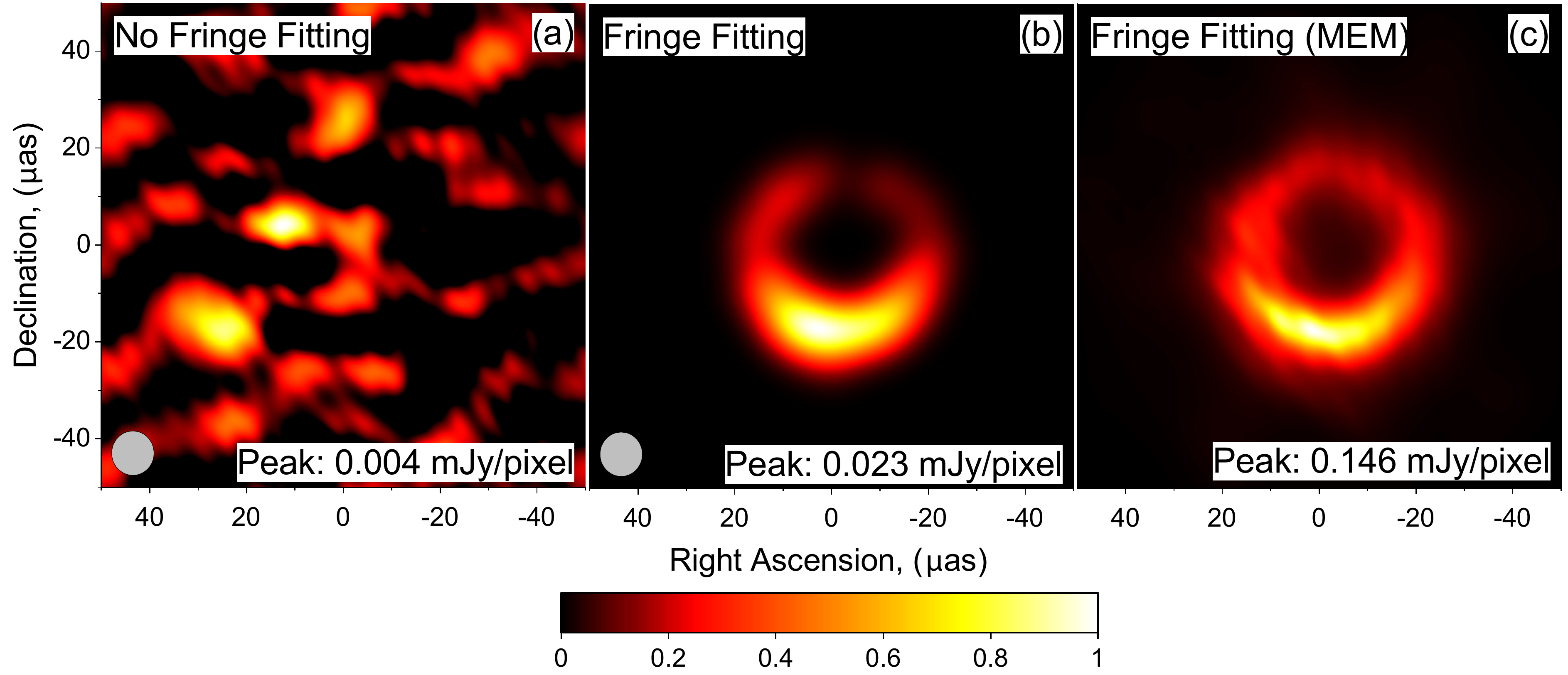}
    \caption{Reconstructed images of M87$^\ast$ for phase noise case (model by Johnson et. al). Left panel shows the reconstructed image without fringe fitting, middle and right panels correspond to the reconstructed images after fringe fitting procedure using CLEAN and MEM methods correspondingly. Estimated beam size for CLEAN image: 7.96 $\times$ 3.53 $\mu$as.}
    \label{fig:fig8}
\end{figure*}

\section{Results} 
In our simulations we used one of the orbits from the family of orbits located around L2 point of the Sun-Earth system from \cite{Rudnitskiy2021}, that is capable of imaging observations of M87$^\ast$ and Sgr~A$^\ast$ sources. 
We performed the corresponding simulations of S-E VLBI observations using several source models. The next step was to evaluate the quality of the obtained images and to compare the estimates to the results of ground-only VLBI and previously obtained values for highly elliptical near-Earth orbits (HEO).

\subsection{M87$^\ast$}

Fig.~\ref{fig:fig4}, Fig.~\ref{fig:fig5} show the resulting images of presented simulations and Fig.~\ref{fig:vis_m87}, Fig.~\ref{fig:fig5b} show the corresponding distribution of visibility amplitude vs. baseline projection for M87$^\ast$ simulations with applied telescope sensitivity (SEFD). The flux level on the presented plots comparing to the estimated $5\sigma$ detection sensitivity of the best baseline in these simulations ALMA+Millimetron of $\approx$20 mJy show good chances to have fringe detection even at long baselines.

Moreover, we have considered several cases of fringe detection for M87$^\ast$. Fig.~\ref{fig:fringe_m87} (top panel, (a)-(d)) show these cases from the best case, when all telescopes gave positive fringe detection (Fig.~\ref{fig:fringe_m87}, (a)) going to the worst case, when only the largest and the most sensitive telescopes gave fringes (Fig.~\ref{fig:fringe_m87}, (d)), i.e. ALMA and LMT. The corresponding resulting images of M87$^\ast$ for these $(u,v)$ coverages are shown in Fig.~\ref{fig:fringe_m87} (bottom panel, (e)-(h)). Despite bad $(u,v)$ coverage in the worst case (Fig.~\ref{fig:fringe_m87} (d)), the results of image reconstruction demonstrate the possibility to still extract any useful information from such observations.

The resulting phase of the visibility model after applying phase errors to the M87$^\ast$ data is shown in Fig. \ref{fig:fig7}. Fig. \ref{fig:fig8} presents the result of M87$^\ast$ image reconstruction  after the global fringe fitting with phase closure. In the fringe fitting procedure the geometric thin ring was used as initial source model. As it can be seen from Fig. \ref{fig:fig4} (e) and Fig. \ref{fig:fig8} (c) the phase calibration using the fringe fitting helps to restore the initial image (see Table~\ref{tab:fidelity_compare}).

Moving forward, to estimate the fidelity for M87$^\ast$ as a reference we used images of model from Fig.~\ref{fig:fig4} (a) and model which is Gaussian blurred with the expected beam size for the Millimetron+EHT combination shown as "Convolved". Result for M87$^\ast$ are shown in Fig. \ref{fig:fig4} and \ref{fig:fig5} and the estimated image quality can be found in Tables \ref{tab:fidelity} and \ref{tab:fidelity_Chernov}.

Comparing the M87$^\ast$ image quality to the results obtained for highly elliptical near-Earth orbit, one can see from Table \ref{tab:fidelity_compare_Earth} that SSIM is 0.345 vs. 0.359 for HEO and L2 (model case) and 0.930 vs 0.928 for HEO and L2 (convolved model case); fidelity estimations are 12.56 vs 7.579 for HEO and L2 (model case) and 22.4 vs 15.805 for HEO and L2 (convolved model case). While, comparing the results of ground only simulations and EHT+Millimetron baselines we have increase in SSIM from 0.064 to 0.359 (model case), from 0.409 to 0.927 (convolved model case) and in fidelity from 3.082 to 7.579 (model case) and from 4.084 to 15.805 (convolved model case).

SANE class models calculated by \citep{Chernov} show similar results comparing EHT-only and EHT+Millimetron baselines. In this case one can observe an increase in SSIM: 0.057 vs. 0.314, 0.086 vs. 0.442 and 0.068 vs. 0.439 (model case) for model \#1, \#2 and \#3 correspondingly. Increase in fidelity for model case:  4.069 vs. 10.095, 5.549 vs. 23.879 and 5.657 vs. 22.384.

Speaking about the sharpness, its values are provided in Table \ref{tab:sharpness}. The example of one-dimensional image profiles for M87 CLEAN images from Fig.~\ref{fig:fig4} ((a), (b) and (d)) that were analyzed for SSIM and sharpness are shown in Fig.~\ref{fig:fig5c}. The lower the RMS value the better the sharpness. In order to avoid the effects of Doppler beaming at high inclination angles onto RMS, the left part of models and obtained image profiles (see Table \ref{tab:sharpness}, last two columns) were analyzed separately. In most cases the EHT+MM synthetic observations show a better correspondence of sharpness comparing to the initial model. 

\subsection{Sgr~A$^\ast$}

Simulated images for Sgr~A$^\ast$ are shown in Fig.~\ref{fig:fig6} and the distribution of visibility amplitude vs. baseline projection shown in Fig.~\ref{fig:vis_sgr}, similarly to M87 taking into account the telescope sensitivity.

And here Sgr~A$^\ast$ shows a diametrically opposite result (image quality estimates are shown in  Table~\ref{tab:fidelity_SgrA}) since there is noise in the image caused by long baseline projections, at which a fine-scale refractive scattering is being resolved. This leads to a drop in fidelity values, compared with what was obtained for the near-Earth orbits from \cite{Andrianov2021}: model \#16 -- 61.39 (HEO) vs. 15.26 (L2); model \#24 -- 49.49 (HEO) vs. 8.9 (L2); model \#31 -- 54.18 (HEO) vs. 10.39 (L2) and model \#39 -- 41.2 (HEO) vs. 7.3 (L2). Reducing the angular resolution (increasing the beam size/shortening the $(u,v)$ tracks), the result is expected to be straightforward comparable to the case of the near-Earth orbit.

As it was shown for M87$^\ast$, in most cases the EHT+MM simulated observations show better sharpness parameters. The only exception here is the case of Sgr~A$^\ast$, namely model \#16 of Sgr~A$^\ast$, where the absence of small-scale details in the model led to the RMS values that are close for both space-ground and ground only cases (see Table~\ref{tab:sharpness}). In general, the sharpness for EHT+MM is $\sim$1.5 times better that for EHT only and, comparing to the near-Earth orbit it is equal or 2-3 times better (except for model \#16), depending on the model number: 0.00069 (HEO) vs. 0.0016 for model \#16, (L2) 0.0045 (HEO) vs. 0.0015 (L2) for model \#24, 0.0019 (HEO) vs. 0.0018 (L2) for model \#31 and 0.0052 (HEO) vs. 0.0029 (L2) for model \#39.

\subsection{General}
As expected EHT+Millimetron observations has superior image quality compared with EHT only. Though SSIM and fidelity for L2 orbit are slightly worse than those for HEO orbit, still these values are $\sim 2-3$ times better than for ground only observations. The image quality figures obtained here for L2 orbit are very similar to the figures obtained by analysis of the near-Earth elliptical orbit \citep{Andrianov2021}. Regarding the imaging methods (CLEAN and MEM), the highest numbers both for fidelity and SSIM method are obtained for MEM image recovery technique both for EHT only and EHT+Millimetron observations.

Considering the possibility of getting a better $(u,v)$ coverage, it should be noted that the selected halo-orbit around the L2 point is not unique and it belongs to the family of stable orbits, while the choice of a specific orbit from this family is dictated by the scientific program of the Millimetron project. Thus, the resulting orbit can be optimized to observe several sources at once with relatively short baseline projection and relatively good $(u,v)$ coverage. Fig. \ref{fig:mol} demonstrates that an orbit optimized for observations of only two sources (M87$^\ast$ and Sgr~A$^\ast$) is also capable of VLBI observations of sources close to the track lines at baseline projections less than 5 Earth diameters. Hence, more sources could be imaged.

\begin{table}
\centering
\caption{Fidelity and SSIM estimate for images obtained by CLEAN and MEM methods for M87$^\ast$ simulations (Fig.~\ref{fig:fig4}). ``Convolved'' corresponds to Gaussian-blurred model with the expected beam size. The column ``Convolved'' means, that the convolved model is compared with the initial model and with itself using SSIM and fidelity measure, the column ``EHT'' corresponds to the SSIM and fidelity measure estimated for EHT-only images and ``EHT+MM'' column corresponds to the SSIM and fidelity measure for EHT+Millimetron images.}
\begin{tabular}{|l|l|l|l|l|}
\hline
\multicolumn{2}{|l|}{CLEAN Method}     & Convolved & EHT       & EHT+MM          \\ \hline
\multirow{2}{*}{SSIM}     & Model      & 0.187      & 0.064     & 0.359           \\
                          & Convolved & 1.000      & 0.409     & 0.927           \\ \hline
\multirow{2}{*}{Fidelity} & Model      & 5.174      & 3.082     & 7.579           \\
                          & Convolved & $\infty$   & 4.084     & 15.805          \\ \hline
\multicolumn{2}{|l|}{MEM Method}       & Convolved & EHT       & EHT+MM          \\ \hline
\multirow{2}{*}{SSIM}     & Model      & 0.187      & 0.114     & 0.177           \\
                          & Convolved & 1.000      & 0.904     & 0.970           \\ \hline
\multirow{2}{*}{Fidelity} & Model      & 5.174      & 3.986     & 5.232           \\
                          & Convolved & $\infty$   & 12.011    & 21.278          \\ \hline
\end{tabular}
\label{tab:fidelity}
\end{table}

\begin{table}
\centering
\caption{Fidelity and SSIM estimate for images obtained by CLEAN method for M87$^\ast$ simulations (Fig.~\ref{fig:fig5}).  ``Convolved'' corresponds to Gaussian-blurred model with the expected beam size. The column ``Convolved'' means, that the convolved model is compared with the initial model and with itself using SSIM and fidelity measure, the column ``EHT'' corresponds to the SSIM and fidelity measure estimated for EHT-only images and ``EHT+MM'' column corresponds to the SSIM and fidelity measure for EHT+Millimetron images.}
\begin{tabular}{|l|l|l|l|l|}
\hline
\multicolumn{2}{|l|}{Model \#1} & Convolved        & EHT       & EHT+MM        \\ \hline
\multirow{2}{*}{SSIM}     & Model      & 0.160      & 0.057     & 0.314           \\
                          & Convolved & 1.000      & 0.381     & 0.946           \\ \hline
\multirow{2}{*}{Fidelity} & Model      & 6.537      & 4.069     & 9.622           \\
                          & Convolved & $\infty$   & 5.356    & 23.519          \\ \hline
\multicolumn{2}{|l|}{Model \#2} & Convolved        & EHT       & EHT+MM        \\ \hline
\multirow{2}{*}{SSIM}     & Model      & 0.248      & 0.086     & 0.442           \\
                          & Convolved & 1.000      & 0.427     & 0.952           \\ \hline
\multirow{2}{*}{Fidelity} & Model      & 6.809      & 4.161     & 10.095           \\
                          & Convolved & $\infty$   & 5.548    & 23.879          \\ \hline
\multicolumn{2}{|l|}{Model \#3} & Convolved        & EHT       & EHT+MM        \\ \hline
\multirow{2}{*}{SSIM}     & Model      & 0.233      & 0.068     & 0.302           \\
                          & Convolved & 1.000      & 0.439     & 0.949           \\ \hline
\multirow{2}{*}{Fidelity} & Model      & 6.983      & 4.162     & 8.650           \\
                          & Convolved & $\infty$   & 5.657    & 22.383          \\ \hline
\end{tabular}
\label{tab:fidelity_Chernov}
\end{table}

\begin{table}
\centering
\caption{Comparison of fidelity and SSIM for M87$^\ast$ simulations with Millimetron on near-Earth orbit \citep[][]{Andrianov2021} and on halo-orbit. ``Convolved'' corresponds to Gaussian-blurred model with the expected beam size. The column ``Convolved'' means, that the convolved model is compared with the initial model and with itself using SSIM and fidelity measure, the column ``EHT'' corresponds to the SSIM and fidelity measure estimated for EHT-only images and ``EHT+MM'' column corresponds to the SSIM and fidelity measure for EHT+Millimetron images.}
\begin{tabular}{|l|l|l|l|}
\hline
\multicolumn{2}{|l|}{}                 & Near-Earth & Halo Orbit \\ \hline
\multirow{2}{*}{SSIM}     & Model      & 0.345     & 0.359       \\
                          & Convolved  & 0.930     & 0.927       \\ \hline
\multirow{2}{*}{Fidelity} & Model      & 12.56     & 7.579       \\
                          & Convolved  & 22.40     & 15.805      \\ \hline
\end{tabular}
\label{tab:fidelity_compare_Earth}
\end{table}

\begin{table}
\centering
\caption{Fidelity and SSIM estimate for images obtained by CLEAN methods for Sgr~A$^\ast$ simulations (Fig.~\ref{fig:fig6}). ``Convolved'' corresponds to Gaussian-blurred model with the expected beam size. The column ``Convolved'' means, that the convolved model is compared with the initial model and with itself using SSIM and fidelity measure, the column ``EHT'' corresponds to the SSIM and fidelity measure estimated for EHT-only images and ``EHT+MM'' column corresponds to the SSIM and fidelity measure for EHT+Millimetron images.}
\begin{tabular}{|l|l|l|l|l|}
\hline
\multicolumn{2}{|l|}{Model \#16}    & Convolved    & EHT       & EHT+MM          \\ \hline
\multirow{2}{*}{SSIM}     & Model      & 0.991      & 0.916     & 0.610           \\
                          & Convolved & 1.000      & 0.949     & 0.567           \\ \hline
\multirow{2}{*}{Fidelity} & Model      & 69.521     & 22.264    & 15.262           \\
                          & Convolved & $\infty$   & 26.872    & 14.239          \\ \hline
\multicolumn{2}{|l|}{Model \#24}       & Convolved & EHT       & EHT+MM          \\ \hline
\multirow{2}{*}{SSIM}     & Model      & 0.994      & 0.944     & 0.441           \\
                          & Convolved & 1.000      & 0.942     & 0.422           \\ \hline
\multirow{2}{*}{Fidelity} & Model      & 64.518     & 18.735    & 8.922           \\
                          & Convolved & $\infty$   & 19.407    & 8.645          \\ \hline
\multicolumn{2}{|l|}{Model \#31}       & Convolved & EHT       & EHT+MM          \\ \hline
\multirow{2}{*}{SSIM}     & Model      & 0.989      & 0.960     & 0.452           \\
                          & Convolved & 1.000      & 0.970     & 0.405           \\ \hline
\multirow{2}{*}{Fidelity} & Model      & 52.949     & 23.248    & 10.389           \\
                          & Convolved & $\infty$   & 27.011    & 9.479          \\ \hline
\multicolumn{2}{|l|}{Model \#39}       & Convolved & EHT       & EHT+MM          \\ \hline
\multirow{2}{*}{SSIM}     & Model      & 0.996      & 0.936     & 0.382           \\
                          & Convolved & 1.000      & 0.935     & 0.372           \\ \hline
\multirow{2}{*}{Fidelity} & Model      & 59.396     & 17.429    & 7.280           \\
                          & Convolved & $\infty$   & 18.214    & 7.207          \\ \hline
\end{tabular}
\label{tab:fidelity_SgrA}
\end{table}

\begin{table}
\centering
\caption{Fidelity and SSIM estimate compare for images of M87$^\ast$ simulations obtained by MEM without phase errors (Fig.~\ref{fig:fig4}(c, e)) and with phase closure after adding errors (Fig.~\ref{fig:fig8}(c)). ``Convolved'' corresponds to Gaussian-blurred model with the expected beam size. The column ``Convolved'' means, that the convolved model is compared with the initial model and with itself using SSIM and fidelity measure, the column ``EHT'' corresponds to the SSIM and fidelity measure estimated for EHT-only images and ``EHT+MM'' column corresponds to the SSIM and fidelity measure for EHT+Millimetron images.}
\begin{tabular}{|l|l|l|l|l|}
\hline
\multicolumn{2}{|l|}{With phase closure} & Convolved & EHT   & EHT+MM          \\ \hline
\multirow{2}{*}{SSIM}     & Model      & 0.187      & 0.114     & 0.114           \\
                          & Convolved & 1.000      & 0.904     & 0.594          \\ \hline
\multirow{2}{*}{Fidelity} & Model      & 5.174      & 3.986     & 4.238           \\
                          & Convolved & $\infty$   & 12.011    & 6.224          \\ \hline
\multicolumn{2}{|l|}{Without phase errors} & Convolved & EHT     & EHT+MM          \\ \hline
\multirow{2}{*}{SSIM}     & Model      & 0.187      & 0.114     & 0.175          \\
                          & Convolved & 1.000      & 0.905     & 0.971           \\ \hline
\multirow{2}{*}{Fidelity} & Model      & 5.174      & 3.994     & 5.203           \\
                          & Convolved & $\infty$   & 12.075    & 21.201          \\ \hline
\end{tabular}
\label{tab:fidelity_compare}
\end{table}

\begin{table*}
\centering
\caption{Calculated sharpness for Sgr~A$^\ast$ and M87$^\ast$. The lower the value the better the sharpness. The columns have the following meaning: EHT-Model corresponds to the sharpness of EHT-only image with the corresponding model subtracted from it, (EHT+MM)-Model corresponds to the sharpness of EHT+Millimetron image the corresponding model subtracted from it; columns with ``left side only'' marks correspond to the estimates of sharpness for the left halves of images.}
\begin{tabular}{l|c|c|c|r}
\hline
 \#             & EHT-Model & (EHT+MM)-Model    & EHT-Model (left side only) & (EHT+MM)-Model (left side only)  \\ 
\hline
Sgr~A$^\ast$    &           &                   &           &                           \\
\hline
16              & 0.0045    & 0.0049            & 0.0016    & 0.0016                    \\
24              & 0.0023    & 0.0018            & 0.0019    & 0.0015                    \\
31              & 0.0050    & 0.0037            & 0.0022    & 0.0018                    \\
39              & 0.0033    & 0.0025            & 0.0029    & 0.0029                    \\
\hline
M87$^\ast$      &           &                   &           &                           \\
\hline
Johnson (Clean) & 0.015     & 0.0091            & 0.01      & 0.001                     \\
Johnson (MEM)   & 0.0098    & 0.0073            & 0.0052    & 0.0038                    \\ 
Chernov \#1     & 0.011     & 0.0074            & 0.0015    & 0.0003                    \\
Chernov \#2     & 0.0098    & 0.007             & 0.001     & 0.0007                    \\
Chernov \#3     & 0.0098    & 0.0061            & 0.0013    & 0.0003                    \\
\hline
\end{tabular}
\label{tab:sharpness}
\end{table*}

\section{Conclusions}
The presented results of SMBH imaging using space-ground interferometer in L2 orbit show a significant improvement in the quality and resolution of the image relative to the observations conducted only with ground-based telescopes. The increase in angular resolution is about $\sim$4 times better, while the image fidelity, SSIM and sharpness values are $\>\sim2$ times better than those for ground-only observations and are comparable to those obtained for highly elliptical orbits in \citep{Andrianov2021}.

Thus, we conclude that the S-E VLBI imaging is fundamentally possible in L2 point orbit and its  capabilities are at least comparable as for HEO. The results of fidelity and SSIM estimations prove that it is possible to obtain images of Sgr~A$^\ast$ and M87$^\ast$ using an optimized orbit in Lagrange L2 point and the analysis of fringe detectability, phase error and sensitivity simulations with several cases of $(u,v)$ coverage degradation confirm this from a technical point of view.

It is important to note, that still there is a capability left for further orbit optimization, so that it would be possible to image other sources (for example Cen~A, \cite{Janssen2021}) and not only Sgr~A* and M87.

\section*{Data availability}
Data underlying related to the results of simulations in this article will be shared on reasonable request to the corresponding author.

\section*{Acknowledgements}
The authors acknowledge Yuri Shchekinov for valuable comments, M.~Mo\'scibrodzka for providing the GRMHD models of Sgr~A$^\ast$ for the described simulations. The authors are grateful to the referee for numerous valuable comments.



\bibliographystyle{mnras}
\bibliography{m_biblio} 



\bsp	
\label{lastpage}
\end{document}